\documentclass[conference]{IEEEconf}

\usepackage{cite}       

\input epsf
\usepackage{graphicx}

\usepackage{cite}
\usepackage{amsmath,amssymb,amsfonts}
\usepackage{algorithmic}
\usepackage[ruled,vlined]{algorithm2e}
\usepackage{graphicx}
\usepackage{textcomp}
\usepackage{xcolor}
\usepackage{footnotebackref}
\usepackage{multirow}
\usepackage{mathtools}
\usepackage{array,booktabs, makecell}
\usepackage{tabularx, threeparttable}

\begin{document}
\title{\textbf{\Large LogGD: Detecting Anomalies from System Logs with Graph Neural Networks\\}}

\author{Yongzheng Xie$^{1}$, Hongyu Zhang$^{2}$, and Muhammad Ali Babar$^{1,*}$\\
	\normalsize $^{1}$The University of Adelaide, Australia\\
	\normalsize $^{2}$The University of Newcastle, Australia\\
	\normalsize yongzheng.xie@adelaide.edu.au, hongyu.zhang@newcastle.edu.au, ali.babar@adelaide.edu.au\\
	\normalsize *corresponding author
}


\maketitle
\begin{abstract}
Log analysis is one of the main techniques engineers use to troubleshoot faults of large-scale software systems. During the past decades, many log analysis approaches have been proposed to detect system anomalies reflected by logs. They usually take log event counts or sequential log events as inputs and utilize machine learning algorithms including deep learning models to detect system anomalies. 
These anomalies are often identified as violations of quantitative relational patterns or sequential patterns of log events in log sequences. However, existing methods fail to leverage the spatial structural relationships among log events, resulting in potential false alarms and unstable performance. In this study, we propose a novel graph-based log anomaly detection method, LogGD, to effectively address the issue by transforming log sequences into graphs. We exploit the powerful capability of Graph Transformer Neural Network, which  combines graph structure and node semantics for log-based anomaly detection. We evaluate the proposed method on four widely-used public log datasets. Experimental results show that LogGD can outperform state-of-the-art quantitative-based and sequence-based methods and achieve stable performance under different window size settings. The results confirm that LogGD is effective in log-based anomaly detection.
\end{abstract}
\IEEEoverridecommandlockouts

%
\IEEEpeerreviewmaketitle

\section{Introduction}
\label{section:introduction}
Modern 
software systems have become increasingly large and complicated\cite{du2017deeplog, zhang2020anomaly, le2021log, yang2021semi}. While these systems provide users rich services, they also bring new security and reliability challenges. One of the challenges is locating system faults and discovering potential issues. 

Log analysis is one of the main techniques engineers use to troubleshoot faults and capture potential risks. When a fault occurs, checking system logs helps detect and locate the fault efficiently. However, with the increase in scale and complexity, manual identification of abnormal logs from massive log data has become infeasible\cite{lou2010mining,du2017deeplog,zhang2020anomaly,yang2021semi}. For example, Google systems generate millions of new log entries every month, meaning tens of terabytes of log data daily \cite{messaoudi2018search,he2017drain}. For such a large amount of data, the cost of manually inspecting logs is unacceptable in practice. Another reason is that a large-scale modern software system, such as an online service system, may comprise hundreds or thousands of machines and software components. Its implementation and maintenance usually rely on the collaboration of dozens or even hundreds of engineers. It is impractical for a single engineer to have all the knowledge of the entire system and distinguish various abnormal logs generated by various software components. Therefore, automated log anomaly detection methods is vital. 

In the past decades, many log-based anomaly detection methods have been proposed. 
Some methods take quantitative log event counts as inputs and utilize traditional Machine Learning (ML) techniques to project the event count vectors into a vector space. Those vectors deviating from the majority (or violating certain invariant relations among the event counts) are classified as anomalies. 
We call such an approach \textit{quantitative-based} approach.
The representative methods of this approach include LR~\cite{bodik2010fingerprinting}, SVM~\cite{liang2007failure}, LogCluster~\cite{lin2016log}, Invariants Mining~\cite{lou2010mining}, ADR~\cite{zhang2020anomaly}, and LogDP~\cite{xie2021logdp}. However, these methods tend to suffer from unstable performance on different datasets since their input only contains quantitative statistics. They fail to capture the rich semantic information embedded in log messages and the sequential relationship between events in a log sequence. 

Recently, deep learning-based methods, such as LogRobust~\cite{zhang2019robust}, CNN~\cite{lu2018detecting}, and NeuralLog~\cite{le2021log}, demonstrate good performance in detecting log anomalies. This class of methods takes sequential log events as input and uses various deep learning models, such as LSTM~\cite{graves2012long}, CNN~\cite{hubel1962receptive} and Transformer~\cite{vaswani2017attention}, to identify anomalies by detecting violations of sequential patterns. We call such an approach \textit{sequence-based} approach. 
Although effective, existing sequence-based methods fail to leverage more informative 
structural relationships among log events, resulting in potential false alarms and unstable performance. 

To address the above issue, this study proposes a novel Graph-based Log anomaly Detection method, namely LogGD. The proposed method first transforms the input log sequences into a graph. It then utilizes the node features (representing log events) and the spatial structure of the graph (representing the relations among log events) 
to detect anomalies, through a customized Graph Transformer Neural Network. The informative spatial structure and the interactions between node features and structural features of the graph
enable the proposed method to better distinguish the anomalies in logs.
Our experimental results on four widely-used public log datasets show that LogGD can outperform the state-of-the-art quantitative-based and sequence-based methods and achieve more stable performance under various window size settings.

Our main contributions are summarized as follows:
\begin{enumerate} 
 \setlength{\itemsep}{-2ex}  
 \setlength{\parskip}{0ex} 
 \setlength{\parsep}{0ex}
\item \textbf{A graph-based log anomaly detection method:} We propose a graph-based anomaly detection method LogGD. The proposed method exploits the spatial structure of log 
graphs and the interactions between node features and structural features for log-based anomaly detection, achieving high accuracy and stable detection performance. 
\hfil\break
\item \textbf{A set of comprehensive experiments:} We compare the proposed methods with five state-of-the-art quantitative-based and sequence-based methods on four widely used real-world datasets. The results confirm the effectiveness of the proposed method.
\end{enumerate}

The rest of the paper is organized as follows. In Section \ref{section:background}, we present the background information and techniques we use in the proposed method. Section \ref{section:methodology} details our proposed methodology. The experimental design and results are described in Section \ref{section:evaluation}. In Section \ref{section:discussion} we discuss why LogGD works and its limitations, as well as threats to validity. We review the related work in Section \ref{section:related_work}), and conclude this work in Section \ref{section:conclusion}.

\section{Background}
\label{section:background}


\subsection{Log Data and Sequences Generation }
\label{subsection:log_parser}

\begin{figure*}[ht] 
    \centering
    \includegraphics[width=0.7\linewidth]{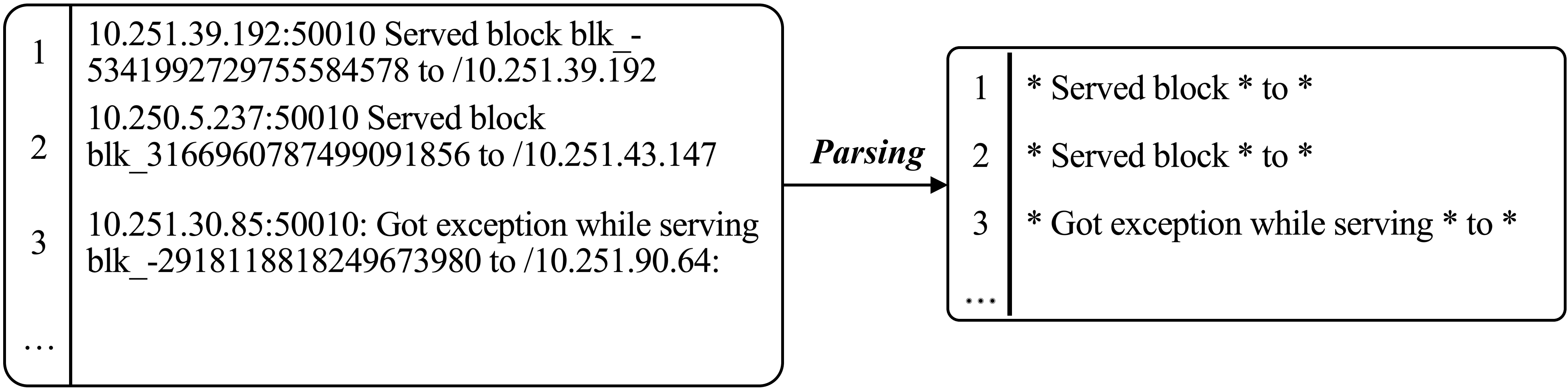}
    \caption{A snippet of HDFS (Hadoop Distributed File System) raw logs and events after parsing}
    \label{fig:LogsSnippet}
\end{figure*}

Logs are usually semi-structured texts which are used to record the status of systems. Each log message comprises a constant part (i.e., log event, also called log template) and a variable part (log parameter). A log parser is a tool that can be used to parse the given log messages into log events. There are many log parsers available~\cite{he2017drain, du2016spell, dai2020logram}. In this study, we choose the state-of-the-art tool, Drain \cite{he2017drain} to complete this task because of its effectiveness, robustness and efficiency that have been validated in~\cite{zhu2019tools}. Figure \ref{fig:LogsSnippet} shows a snippet of raw logs and the results after they are parsed.

Once the structured log events are ready for use, they need to be further grouped into log sequences (i.e., series of log events that record specific execution flows) according to sessions or predefined windows. Session-based log partition often utilizes certain log identifiers to generate log sequences. Previous studies have \cite{he2016experience,le2022log} shown that the methods working on session-based data can achieve better performance than on data grouped by predefined windows. However, many datasets cannot find such identifiers to group log data. In such cases, predefined windows are a common choice for log partitioning, where two strategies are available, i.e., fixed and sliding windows. The fixed window strategy uses a predefined window size, e.g., 100 logs or 20 logs, to produce log sequences with a fixed number of events. In contrast, the sliding windows strategy generates log sequences with overlapping events between two consecutive windows, where window size and step size are two attributes used. An example is 20 logs windows(window size) sliding every 1 log(step size). At this point, the resulting log sequence can be used for graph data generation. 

\subsection{Graph Neural Networks}
\label{subsection:Graph_Neural_Networks}

Recently, Graph Neural Networks(GNNs) have attracted great interest from researchers and practitioners as GNN-based methods have achieved excellent performance on many tasks such as drug discovery~\cite{jiang2021could}, social network analysis~\cite{min2021stgsn}, network intrusion detection~\cite{zhou2021hierarchical}, and so on. We introduce our notation and then briefly recap the preliminaries in graph neural networks.
 
 \textbf{Notation:} Let G = ($V$, $E$, $X_v$, $F_e$) denote a directed graph,  where $V=\{v_i | i\in \mathbb{N}_0\}$ represents a set of nodes and $E = \{e_{{v_i}{v_j}} |  v_i, v_j \in V\}$ refers to a set of directed edges that flows from node $v_i$ to node $v_j$, ${X_v} \in \mathbb{R}^{|V| \times d_v}$ denotes the node features with dimension $d_v$ and ${F_e} \in \mathbb{R}^{|E| \times d_e}$ represent the edge features with dimension $d_e$. We use $N_{v_i}$ to denote the set of neighbors of node $v_i \in {V}$, i.e., $N_{v_i} =\{v_j| e_{{v_i}{v_j}} \in E\}$. 
 
 \textbf{Graph Neural Networks:} Unlike sequence models on text data ((such as LSTM, GRU and Transformer) and convolutional models on image data ((such as CNN), graph neural networks work on graph data with irregular structures rather than sequential or grid structures. They combine the graph structure and node features to learn high-level representation vectors of the nodes in the graph or a representation vector of the whole graph for node classification, link/edge prediction or graph classification. In this work, we focus on the task of graph classification, i.e., predicting the label for a given graph.
 
GNNs typically employ a neighborhood aggregation strategy~\cite{dwivedi2021graph,bruel2022rewiring}, where the representations of nodes in the graph are iteratively updated by aggregating representations of their neighbors. Ultimately, a node's high-level representation captures structural attributes within its L-hop network neighborhood. Formally, the $l$-th layer representation of node $v_i$ can be formulated as:
 \begin{equation}
 \begin{aligned}
 \label{eq:1}
  {x^{l}_{v_i}} = {\boldsymbol{\Phi}(x^{l-1}_{v_i}, \boldsymbol{\Psi}_{v_j \in N_{v_i} }(x^{l-1}_{v_j}, f_{e_{{v_i}{v_j}}}))} 
  \end{aligned}
 \end{equation}
 where $f_{e_{{v_i}{v_j}}} \in F_e$ denotes the feature vectors of edges $\{e_{{v_i}{v_j}}|v_j \in N_{v_i}\}$ and $x^0_{v_i} \in X_v$ is the initial node feature vector of a graph, and $N_{v_i}$ denotes the neighbourhood of node $v_i$, $\boldsymbol{\Psi}$ and $\boldsymbol{\Phi}$ represent the abstract functions of the graph encoder layer for information gathering from neighbors as well as information aggregation into the node representation, respectively.
 
 For graph classification tasks, the derived node high-level representations from Equation \ref{eq:1} need to be further aggregated into a graph-level representation through a function named $READOUT$, which usually is performed at the final layer of graph encoder as follows:
 \begin{equation}
  \begin{aligned}
  \label{eq:2}
   h_{g} = {\boldsymbol{\Omega}(\omega;X^{l}) = READOUT(x^{l}_{v_i}|v_i \in V}) 
  \end{aligned}
 \end{equation}
 where $h_g$ denotes the representation of the given graph $g$, and $X^l$ represents the node representation matrix. $READOUT$ is a parameterized abstract function $\boldsymbol{\Omega}(\cdot)$ with parameters $\omega$, which can be implemented as any aggregation function, such as sum, max, mean-pooling or more complex approach in real applications.  
 
\subsection{Graph Transformer Networks}
\label{subsection:Graph_Transformer_Networks}
The Transformer architecture originates from the field of Natural Language Processing(NLP)~\cite{vaswani2017attention}. Due to its excellent performance on various language tasks, it has been generalized to graph-based models~\cite{dwivedi2020generalization,ying2021transformers,park2022grpe}, i.e., Graph Transformer(GT) model. A graph transformer block comprises two key components: a self-attention module and a position-wise feed-forward network(FFN). The self-attention module first projects the initial node features $X \in \mathbb{R}^{|V| \times d_v}$ into query(Q), key(K), value(V) matrices through independent linear transformations, where $W^{query} \in \mathbb{R}^{{d_v} \times {d_z}}$,  $W^{key} \in \mathbb{R}^{{d_v} \times {d_z}}$, and $W^{value} \in \mathbb{R}^{{d_v} \times {d_z}}$ denote learnable parameters, $d_z$ denotes the output dimension of linear transformation. 
\begin{equation}
 \begin{aligned}
 \label{eq:transformer_1}
  {Q = X \cdot W^{query}, \quad K = X \cdot W^{key}, \quad and \quad V = X \cdot W^{value}}
  \end{aligned}
 \end{equation}
then the attention coefficient matrix can be obtained through a scaled dot production between queries and keys.
\begin{equation}
 \begin{aligned}
 \label{eq:transformer_2}
  {a_{ij} = \frac{{q}_{i} \cdot k_{j}}{\sqrt{d_z}} \quad and \quad \hat{a_{ij}} = \frac{\exp(a_{ij})}{\sum_{k=1}^{N}{\exp(a_{ik})}}}
  \end{aligned}
 \end{equation}
Next, the self-attention module outputs the next hidden feature by applying weighted summation on the values.
\begin{equation}
 \begin{aligned}
 \label{eq:transformer_3}
  {h^{l}_{i} = {\sum_{j=1}^{N}{\hat{a_{ij}}v_j}}}
  \end{aligned}
 \end{equation}
In order to improve the stability of the model, the multi-head mechanism is often adopted in the self-attention module. After that, the output of the self-attention is followed by a residual connection and a feed-forward network and ultimately provides node-level representations of the graph. Finally, a $READOUT$ function in the equation~\ref{eq:2} is applied to the final layer output of the graph transformer model to obtain the graph representation. 

\section{Proposed Method}
\label{section:methodology}

\subsection{Overview}

\begin{figure*}[ht]
  \centering
  \includegraphics[width=0.7\linewidth]{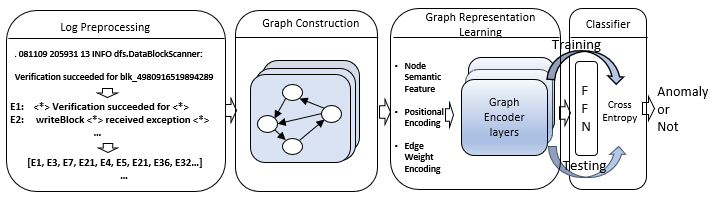}
  \caption{The overview of the proposed method.}  
  \label{fig:framework_overview}
\end{figure*}

 The proposed method, LogGD, is a graph-based log anomaly detection method that consists of three components: graph construction, graph representation learning, and graph classification. The input is the 
 log sequences generated in section~\ref{subsection:log_parser}, and the output is whether a given log sequence is anomalous or not. LogGD starts by transforming the given log sequences into a graph. 
 The node features contain the semantic information of log events, and the edges include the connectivity and weights of pairs of nodes. Then, the resulting graph data is fed into a GNN model 
 to learn the patterns of normal and abnormal graphs in the training phase. During the testing (inference) phase,  the representation of a given graph obtained through the same process is 
 classified as anomalous or non-anomalous. 

\subsection{Graph Construction}
\label{subsection:graph_construction}

First, each log sequence derived from section~\ref{subsection:log_parser} is transformed into a directed graph, denoted as ${G}$ = (${V}$, ${E}$, ${X_v}$, ${F_w}$), where ${V}$ represents the set of nodes $v_i$,  corresponding to the log events of the log dataset and $E$ refers to the set of edges $e_{{v_i}{v_j}}$, that is, a node pair where event $v_i$ is immediately followed by event $v_j$ in the sequence. ${X_v} \in \mathbb{R}^{|V| \times d}$ denotes the node features corresponding to the semantic vectors of log events generated by some NLP technique. ${F_w} \in \mathbb{R}^{{|E|} \times 1}$, i.e., the set of edge weights, indicating the occurrence frequency of the edges $e_{{v_i}{v_j}}$ in a sequence.  It is worth noting that a self-loop edge is always added for the initial event since there is no preceding event before the initial event. In addition, the node set $V$ and their corresponding initial node features ${X_v}$ are shared across the graphs transformed from the same log dataset. Through the previous steps, we construct the graphs from a given set of log sequences.

\begin{figure*}[ht]
  \centering
  \includegraphics[width=0.7\textwidth]{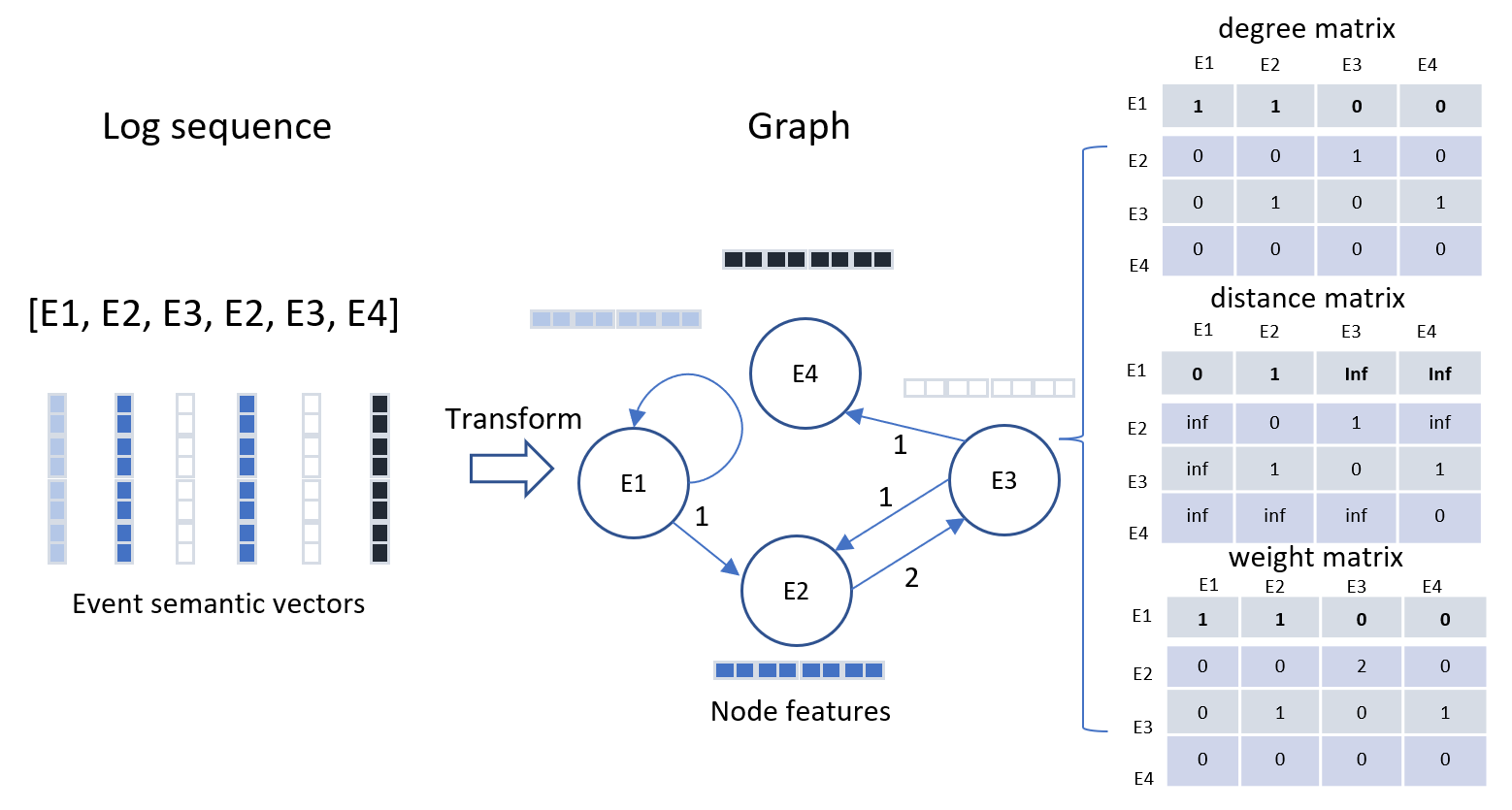}
  \caption{The diagram of the graph construction from a log sequence.} 
  \label{fig:graph_construction}
\end{figure*}

Fig.~\ref{fig:graph_construction} is an example that transforms a log sequence [E1, E2, E3, E2, E3, E4] with the event semantic vectors into a graph consisting of node features and graph structure attributes. From the figure, we can see that a graph can provide richer spatial structure attributes than a sequence of log events. The spatial structure of a graph includes the node-centered local structure represented by the degree matrix, the global structure of node locations encoded by a distance matrix, and the quantitative connections between nodes represented by the weight matrix. Later, We will see that the combination of the spatial structure attributes will benefit the graph-based methods to generate more expressive representations of sequences, thereby improving detection accuracy and stability. An important point to note is that a graph contains only nodes with no duplicates, unlike sequences that allow duplicate log events. Taking the sequence above as an example, although events E2 and E3 appear twice in the sequence, node E2 and node E3 in the transformed graph exist as a single node, respectively. The occurrence frequency information of the nodes is reflected in the corresponding edge weights, ensuring no information is lost in the resulting graph.

\subsection{Graph Representation Learning}
\label{subsection:representation_generation}

Graph representation learning is the process of learning an expressive low-dimensional representation incorporating node features and spatial structure attributes of a given graph, which is crucial in graph classification tasks. In order to make the trained model more discriminative towards normal and abnormal graphs, we need to carefully design the features that will participate in generating the graph representation.

\textbf{Semantic-Aware Node Embedding:}
Each node in a graph represents a unique log event derived from the log parsing process. As many prior studies \cite{zhang2019robust,chen2021experience,le2021log,le2022log} show, the semantic information embedded in the log messages can have a significant impact on the performance of subsequent log anomaly detection. To extract semantics from text data, there are many NLP models available, such as Word2Vec~\cite{nguyen2016integrating}, Glove~\cite{pennington2014glove}, FastText~\cite{joulin2016fasttext}, and BERT model~\cite{devlin2018bert}. In this study, we utilize the BERT model to extract semantics embedded in the log messages because it has been proved to better capture and learn the similarity and dissimilarity across log messages based on the position and context of words\cite{huang2020hitanomaly,le2021log}. We follow \cite{le2021log} to tokenize each template into a set of words and subwords and employ the feature extraction function of pre-trained BERT \footnote{\url{https://github.com/google-research/bert}} to obtain the semantic information of each log event. Finally, each log event is encoded into a vector representation with a fixed dimension. In our experiments, this fixed dimension is 768.

\textbf{Structure-Aware Encoding for Graph:}
For log sequences, the sequential relationship between log events is often an important indicator of normality or abnormality. The sequential relation between log events reflects the positional structure of log events in the sequence. Sequence-based methods either implicitly exploit the positional structure of log events by sequentially processing a given sequence (e.g., LogRobust~\cite{zhang2019robust})  
or employ explicit positional encoding to enhance the sequence representations (e.g., NeuralLog~\cite{le2021log}). 
Unlike sequence data, graphs typically do not have such a node sequential structure due to the invariance of graphs to node permutation. Inspired by \cite{park2022grpe,ying2021transformers}, in this study, we utilize three structural attributes of graph, the degree matrix, the distance matrix, and the edge weight matrix, to generate structure-aware encoding for a graph to enhance the discriminatory of graph representation. Intuitively, the in-degree and out-degree of a node reflect the local topology of a node. It not only represents the importance of a node in the graph, but also reflects the similarity between nodes, 
which can be used to complement the semantic similarity between nodes. In addition, the shortest path distance matrix of a graph reflects the global spatial structure of a graph, while the edge weight matrix incorporates the quantitative relation of connections between nodes. Thus, the degree matrix, distance matrix, and edge weight matrix reflect different aspects of a graph and theoretically complement each other. When combined with node features to generate a representation for a graph, they will help enhance the representation of a given graph for graph classification. 

As an example, let us assume [E1, E2, E3, E4, E5, E2, E6] as a normal sequence, while [E1, E2, E3, E4, E5, E2, E2, E6] is abnormal because an additional event, E2, occurs in the penultimate position. Although the degree and distance matrix of the graph cannot catch the anomalous information, the change can be reflected in the edge weight matrix. 
As another example, suppose that [E1, E2, E3, E4, E7, E2, E6] is an anomaly because E7 replaces event E5. In this case, the changes will be reflected in the entries of all the matrices of the graph, including the distance matrix, the degree matrix, and the edge weight matrix. Similarly, whether the sequential relationship between a pair of events changes, or one event is inserted or substituted for another, the anomaly can always be reflected in an aspect of graph structure-aware coding. Therefore, structure-aware encoding can be expected to help enhance the representation of a given graph for graph classification.

\textbf{Graph Representation Learning:}
The informative graph spatial structure attributes and the node features require GNN models to digest and produce the high-level representation for a given graph. There are many GNNs models available, such as GCN~\cite{kipf2016semi}, GCNII~\cite{chen2020simple}, GAT~\cite{velivckovic2017graph}, GATv2~\cite{brody2021attentive}, GIN~\cite{xu2018powerful}, GINE~\cite{hu2019strategies}, and Graph Transformer Network(TransformerConv)~\cite{shi2020masked}. In this study, we choose Graph Transformer Network architecture for graph representation learning because it not only overcomes the limitations of under-reaching and overs-quashing in the stacked layers of message-passing GNNs~\cite{bruel2022rewiring}, but also exhibits better performance than GAT models in graph classification by exploiting positional/structural encoding~\cite{park2022grpe,ying2021transformers}. The input to our graph representation learning models is graphs derived from \ref{subsection:graph_construction}. Structure-aware encoding of graphs is used for graph representation learning, including degree matrices, distance matrices, and edge weight matrices.

First, for the degree matrix embeddings, inspired by~\cite{ying2021transformers}, we add them to the node features. By doing this, the model can capture the node similarity represented by the node-centered local topology and the semantic correlations embedded in the node features through an attention mechanism.

\begin{equation}
\begin{aligned}
\label{eq:degree_matrix}
{x^{0}_i = x_{v_i} + z_{indeg(v_i)} + z_{outdeg(v_i)} }
\end{aligned}
\end{equation}
where $x_{v_i} \in \mathbb{R}^{d_v}$ denotes the feature of node ${v_i}$, $z_{indeg(v_i)} \quad  and \quad z_{outdeg(v_i)} \in \mathbb{R}^{d_v}$ are learnable embedding vectors corresponding to the in-degrees and out-degrees of node $v_i$, respectively.

Second, we encode quantitative connection relationships between nodes and incorporate edge weight information into Q and V by element-wise multiplication:

\begin{equation}
 \begin{aligned}
  {Q^0 = w \circ Q, \quad K^0 = K, \quad and \quad V^0 = w \circ V}
  \end{aligned}
 \end{equation}
where $w$ denotes the summation of the edge weight matrix along the row dimension, $Q$, $K$, and $V$ represents the output of the projection from the node features of the graph by a linear transformation in the equation~\ref{eq:transformer_1}.

In the third step, to encode a graph's global structure attribute, we do not directly add the relative path distance scalar to the attention coefficients. Instead, we adopt the practice described in ~\cite{park2022grpe} to define the distance embeddings $D \in \mathbb{R}^{L \times d_z } $, which represents the relative path distance with the maximum length $L$ between nodes in a graph. The distance embeddings are shared throughout all layers. The global structure is then encoded as a spatial bias term $b^{spatial}_{ij}$, computed as the sum of dot products between the node feature and the distance embedding. Notably, the sum of dot products between the node feature and the distance embedding reflects the interaction between node features and graph structure. As such, the proposed spatial bias term enables the trained model to distinguish different structures even when two nodes have the same distance. 

\begin{equation}
\begin{aligned}
\label{eq:3}
{b^{spatial}_{ij} = q_i \cdot D_{\psi_{(ij)}} + k_j \cdot D_{\psi_{(ij)}}}
\end{aligned}
\end{equation}

Then, the spatial bias term $b^{spatial}_{ij}$ is added to the scaled dot product attention coefficients matrix to encode the global structural attribute.

\begin{equation}
\begin{aligned}
\label{eq:5}
{a_{ij} = \frac{q_i \cdot k_j + b^{spatial}_{ij}}{\sqrt{d_z}}}
\end{aligned}
\end{equation}

Finally, the node features are encoded as hidden features by weighted summation of the value and spatial bias terms with the attention coefficient matrix:

\begin{equation}
\begin{aligned}
\label{eq:6}
{z_{i} = \sum^N_{j=1}{\hat{a_{ij}}}({v_j + D_{\psi_{(ij)}}})}
\end{aligned}
\end{equation}

Our method encodes both the node-wise information (attention coefficient) and the interaction-wise information between the node and structure of a graph into the hidden features of value, which is different from those methods encoding only node-wise information. Thus, when the attention weight $\hat{a}$ is applied equally to all channels, our graph-encoded value enriches the feature of each channel. 

	\begin{table*}[tb]
		\renewcommand{\arraystretch}{0.9}
		\centering
		\caption{Overview of datasets used in the experiments.}
		\resizebox{0.9\textwidth}{!}{
		\begin{threeparttable}
		\begin{tabular}{ lcl cccccccr}
			\toprule
			\multirow{2}{*}{Datasets} &  \multirow{2}{*}{\#Nodes}  & \multirow{2}{*}{Window} & \multicolumn{4}{c}{Training Set(80\%)} & \multicolumn{4}{c}{Testing Set(20\%)}\\
			\cmidrule{4-11} 
			& & &\#Graphs  &\#Nodes &\#Anom.  & \%Anom.  &\#Graphs  &\#Nodes &\#Anom.  &\%Anom.  \\
			\toprule
			HDFS& 48 & session &46,0048 &48 &13,470 &2.93\% &115,013 &43 &3,368 &2.93\% \\
			\midrule 
			&        & 100 logs &37,708 &980 &4,009 &10.63\% &9,427 &1,063 &817 &8.66\% \\
			\cmidrule{3-11}
			BGL& 1847& 60 logs&62,847 &980 &6,307 &10.04\% &15,712 &1,063 &1,194 &7.60\%\\
			\cmidrule{3-11}
			& & 20 logs&188,540 &980 &17,252 &9.15\% & 47,135 &1,063 &3,006 &6.38\%\\
			\midrule
			& & 100 logs&63,867 &1,209 &20,195 &31.62\% & 15,967 &844 &399 &2.50\% \\
			\cmidrule{3-11}
			Spirit& 1,229& 60 logs&106,445 &1,209 &30,882 &29.01\% & 26,612 &844 &410 &1.54\% \\
			\cmidrule{3-11}
			& & 20 logs&319,334 &1,209 &81,550 &25.54\% & 79,834 &844 &438 &0.55\%\\
			\midrule
			& & 100 logs&79,674 &3,779 &816 &1.02\% & 19,919 &1,923 &27 &0.14\% \\
			\cmidrule{3-11}
			TDB& 4,992& 60 logs&132,789 &3,779 &985 &0.74\% & 33,198 &1,923 &34 &0.10\% \\
			\cmidrule{3-11}
			& & 20 logs&398,367 &3,779 &1,394 &0.35\% & 99,592 &1,923 &48 &0.01\%\\

			\bottomrule
		\end{tabular}
		\begin{tablenotes} 
		\item \#Nodes: number of unique events; \#Graphs: number of sequences; \#Anom.: number of anomalies; \%Anom.: percentage of anomalies.
		\end{tablenotes}
		\end{threeparttable} 
		}
		\label{tab:tab1}
		\vspace{-10pt}
	\end{table*}

\subsection{Anomaly Detection through Graph Classification}
\label{subsection:anomaly_detection}

To implement the classification task, the output graph representation of the graph encoder layer is directly connected to a feed-forward network with layer normalization(LN)~\cite{ba2016layer}, which contains three fully connected layers with Gaussian Error Linear Units(GELU)~\cite{hendrycks2016gaussian} as the activation function. The sum and maximum values of the output node representations are concatenated as the $READOUT$ function for the graph representation. Then, the cross-entropy is used as the loss function, and the class probabilities of normality and abnormality for the given log sequences are calculated using the $softmax$ function: 

 \begin{equation}
  \begin{aligned}
  \label{eq:7}
   h_{g} = concat(\sum_{v_i \in V}{x^{l}_{v_i}}, \max_{v_i \in V}(x^{l}_{v_i})) 
  \end{aligned}
 \end{equation}

\begin{equation}
\begin{aligned}
\label{eq:8}
{\hat{y_g} = softmax(FFN(h_g))}
\end{aligned}
\end{equation}

In this way, we train a GNN-based model for log-based anomaly detection. When a set of new log messages are provided, they are first preprocessed. Then the new log messages are transformed into semantic vectors as node features, and the sequences are converted to graphs. Afterward, the resulting graph data is fed into the trained model. Finally, the GNN-based model can predict whether the given graph is anomalous or not.

\section{Evaluation}
\label{section:evaluation}

\subsection{Datasets}

In our experiments, four public log datasets, HDFS, BGL, Spirit and Thunderbird, are used to evaluate the proposed approach and the relevant baseline methods. The datasets are widely used in log analysis research~\cite{le2021log,zhang2020anomaly,du2017deeplog,lou2010mining,xu2009largescale,landauer2022deep} because all of them come from real-world datasets and are labeled either manually by system administrators or through alert tags automatically generated by their systems. We obtained all the log datasets from the publicly available websites.\footnote{\url{https://github.com/logpai/loghub} and \url{https://www.usenix.org/cfdr-data}}. Further details about the four datasets are described as follows.

\textbf{HDFS} dataset is generated by running Hadoop-based map-reduce jobs on more than 200 Amazon’s EC2 nodes. 
Among the 11,197,954 log entries collected, approximately 2.9\% of them are abnormal. 

\textbf{BGL} dataset is an open dataset of logs collected from a BlueGene/L supercomputer system at Lawrence Livermore National Labs (LLNL) in Livermore, California, with 131,072 processors and 32,768GB memory. 

\textbf{Spirit} dataset is a high-performance cluster that is installed in the Sandia National Labs(SNL). The Spirit dataset was collected from the systems with 1,028 processors and 1,024GB memory. In this study, we utilize 1 gigabyte continuous log lines from Spirit data set for computation-time purposes. 

\textbf{Thunderbird (TDB)} 
dataset is also from the supercomputer at Sandia National Labs (SNL). 
The dataset is a large dataset of more than 200 million log messages. We only leverage 10 million continuous log lines for computation-time purposes.

The details of the log datasets in our experiments are summarized in Table~\ref{tab:tab1}. From the table, we can see that these datasets exhibit diversity in node size and anomaly rate, which can better validate the generalization of the evaluation method.

\subsection{Implementation and Experimental Setting}
We implemented LogGD and its variants based on Python 3.8.5, PyTorch 1.11.0 and PyG 2.04, respectively. For the GCNII, GINE, GATv2 and TransformerConv models, we utilized the corresponding modules from PyG with default parameters setting. In our experiments, we set the graph encoder layer size of LogGD as 1. The size of the feed-forward network that takes the output of the encoder layer is 1024. LogGD is trained using AdamW optimizer. The linear learning rate decay is used, and the learning rate starts from $3\ast10^{- 4}$ and ends at $1\ast10^{- 9}$. We set the mini-batch size and the dropout rate to 64 and 0.3, respectively. We use the cross-entropy as the loss function.The model trains for up to 100 epochs and performs early stopping for 20 consecutive iterations without loss improvement.

Regarding the baseline approaches used for comparison, we adopt the implementations in the studies\cite{he2016experience, chen2021experience, le2021log}. We use the parameters set in their implementation. In the experiments, we ran each method three times at each window setting on the four datasets and averaged them as the final result to report.

We conduct all the experiments on a Linux server with AMD Ryzen 3.5GHz CPU, 96GB memory, RTX2080Ti with 11GB GPU memory and the operating system version is Ubuntu 20.04. 

\subsection{Compared Methods}
To evaluate the effectiveness of the proposed method, we compare LogGD with five state-of-the-art existing supervised log anomaly detection methods on the aforementioned public log datasets. Specifically, there are two quantitative-based methods, LR\cite{bodik2010fingerprinting} and SVM\cite{liang2007failure}, and three sequence-based methods, CNN\cite{lu2018detecting}, LogRobust\cite{zhang2019robust} and NeuralLog\cite{le2021log}.

We did not directly compare the proposed method with another state-of-the-art graph-based method GLAD-PAW~\cite{wan2021glad} because GLAD-PAW is a semi-supervised method. However, we implemented a corresponding supervised method using the GAT model and compare it with our proposed method in the subsequent experiments.

\subsection{Evaluation Metrics}

To evaluate the effectiveness of the approaches, we utilize precision/recall/F1 score as the metrics, which are widely used in many studies\cite{he2016experience, chen2021experience, le2021log}. Specifically, the metrics are calculated as follows: 
\hfill

\begin{itemize}
\item \textbf{Precision:} the percentage of correctly detected abnormal log sequences amongst all detected abnormal log sequences by the model. $Precision = \frac{TP}{TP+FP}$

\item \textbf{Recall:} the percentage of log sequences that are correctly identified as anomalies over all real anomalies. $Recall = \frac{TP}{TP+FN}$ 

\item \textbf{F1:} the harmonic mean of Precision and Recall. 
$F1 = \frac{2*Precision*Recall}{Precision+Recall}$
\end{itemize}

\subsection{Research Questions}
Our experiments are designed to answer the following research questions:

\begin{itemize}

\item \textbf{RQ1. How effective is the proposed graph-based approach for log anomaly detection?}

\item \textbf{RQ2. Can the proposed approach work stably under various window settings?}

\item \textbf{RQ3. How does LogGD perform with other GNN models?}

\item \textbf{RQ4. How do the specific structural features and the interaction affect the performance of LogGD?}

\end{itemize}

\subsection{Results and Analysis}
\hfill

\textbf{RQ1. How effective is the proposed graph-based approach for log anomaly detection?}

	\begin{table*}[tb]
        \renewcommand{\arraystretch}{0.9} 
    	\centering
    	\caption{Experimental results of LogGD and benchmark methods.}
        \tiny
    	\label{tab:results}%
    	\resizebox{0.8\textwidth}{!}{ 
    		\begin{tabular}{lc cccccc}
    			\toprule
    			\textbf{Dataset} & \textbf{Metrics} & \textbf{LogGD} & \textbf{LR}   & \textbf{SVM} & \textbf{LogRobust} & \textbf{CNN} & \textbf{NeuralLog} \\
    			\toprule
    			\multirow{3}[0]{*}{ HDFS } & F1    & \textbf{0.9877} & 0.9616 & 0.8330 & 0.9819 & 0.9872 & 0.9827 \\
    			& Precision & 0.9774 & 0.9603 & 0.9519 & 0.9688 & \textbf{0.9852} & 0.9627  \\
    			& Recall & \textbf{0.9982} & 0.9629 & 0.7405 & 0.9954 & 0.9891 & 0.9956  \\
    			\cmidrule{2-8}
                \multirow{3}[0]{*}{	BGL } & F1    & \textbf{0.9719} & 0.2799 & 0.4558 & 0.9402 & 0.9140 & 0.9535  \\
    			& Precision & \textbf{0.9708} & 0.1684 & 0.8190 & 0.9229 & 0.8669 & 0.9586  \\
    			& Recall & \textbf{0.9731} & 0.8286 & 0.3158 & 0.9596 & 0.9702 & 0.9484  \\
    			\cmidrule{2-8}
    			\multirow{3}[0]{*}{	Spirit } & F1  & \textbf{0.9789} & 0.9652 & 0.9736 & 0.9757 & 0.9652 & 0.9510  \\
    			& Precision & 0.9889 & 0.9580 & 0.9773 & \textbf{0.9957} & 0.9740 & 0.9694  \\
    			& Recall & 0.9691 & \textbf{0.9724} & 0.9699 & 0.9566 & 0.9566 & 0.9349  \\
    			\cmidrule{2-8}
    			\multirow{3}[0]{*}{ TDB } & F1 & \textbf{0.9284} & 0.4651 & 0.7797 & 0.4043 & 0.5533 & 0.7704  \\
    			& Precision &\textbf{0.9772} & 0.3390 & 0.7188 & 0.4329 & 0.5405 & 0.9683 \\
    			& Recall & \textbf{0.8889} & 0.7407 & 0.8519 & 0.4198 & 0.5802 & 0.6437  \\
    			\bottomrule
    		\end{tabular}%
    	}
    	\vspace{-10pt}
     \end{table*}%

In this experiment, we aim to evaluate the effectiveness of LogGD on the four aforementioned public log datasets. To generate the log sequences, we use session window on the HDFS dataset to group log messages by the same block ID, as the data is labeled by blocks. Then, 80\% of log sequences are randomly selected for training, and the rest of the dataset is used for testing. For the BGL, Spirit and Thunderbird datasets, we keep the chronological order of the dataset and leverage the first 80\% log messages as the training set and the rest 20\% as the testing set, which aims at following the real scenarios and ensures new log events that are unseen in the training set appearing in the testing set. We group log sequences on the BGL, Spirit and Thunderbird datasets by fixed-window rather than by session or sliding-window because there is no universal identifier available for session grouping, and the fixed-window grouping strategy is more storage efficient than sliding-window grouping strategy. In this experiment, the fixed-window size of the input data is set to 100 logs. We present the results for other window-size settings in the subsequent research question. In addition, we utilize oversampling technique to address the imbalance of the training data. If the anomaly rate of the training data is less than 30\%, we oversample it to 30\%; otherwise, we do not use oversampling. To make a fair comparison, all the deep learning-based methods, LogRobust, CNN, NeuralLog and LogGD all take the semantic vectors of log messages that are generated by Bert model as input. In addition, the 10\% of the training set on each dataset is used as the validation set to decide when to early stop the training.

The experimental results are shown in table \ref{tab:results}. It can be seen that LogGD outperforms all comparison methods on four datasets under a fixed window setting of 100 logs with an improvement of F1 score by 0.5\% on HDFS and 1.9\% on BGL, 0.5\% on Spirit, 19.1\% on Thunderbird dataset compared to the second best baseline method. Meanwhile, from the table, we can also see that all the baseline methods but LogGD perform poorly on the Thunderbird datasets because the Thunderbird dataset has highly imbalanced data with a small percentage of anomalies(1.02\% and 0.14\% in the training and test datasets, respectively). Although oversampling has been applied to data preprocessing in experiments, limited by too few anomalies, this still does not prevent poor performance of all baseline methods. In contrast, LogGD performs much better, albeit with a slight drop in performance on the Thunderbird dataset. This may be attributed to the advantage that our method can capturing additional graph structure information to help distinguish the difference between normal and abnormal graphs. Furthermore, although both NeuralLog~\cite{le2021log} and LogGD belong to the transformer architecture, the utilization of graph structure information rather than sequential position encoding can still enable our proposed graph-based method to achieve better performance.

\textbf{RQ2. Can the proposed approach work stably under various window settings?}

This experiment aims to investigate whether LogGD can work stably under various window settings. We conduct the experiments only on the BGL, Spirit and Thunderbird datasets because HDFS only has a session window setting. The experiment is implemented under three window size settings, i.e., 100logs, 60logs, and 20logs. Some larger window size settings were not chosen in this experiment because they are generally unlikely to be adopted in real scenarios due to potential delays in fault discovery. 

\begin{figure*}[!htb]
  \includegraphics[width=0.3\linewidth]{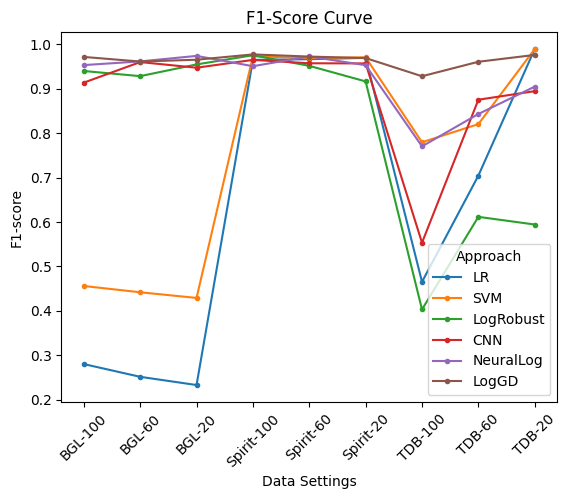} \quad
  \includegraphics[width=0.3\linewidth]{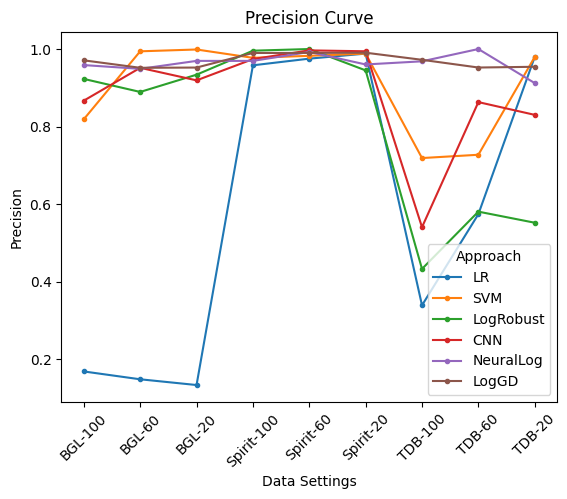} \quad
  \includegraphics[width=0.3\linewidth]{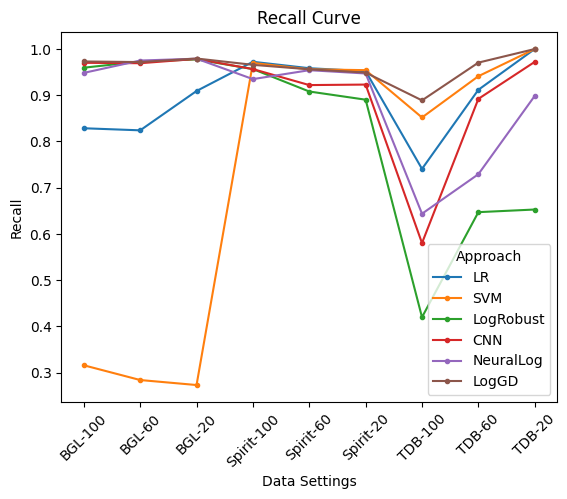}
\caption{Performance Comparison under Different Window Setting between Baseline Methods and LogGD}\label{fig:window_comparision}
\end{figure*}

The comparison results between baseline methods and LogGD are shown in Fig~\ref{fig:window_comparision}. From the figures, we can see that LogGD works more stably on the three window size settings of each dataset compared to the quantitative-based and sequence-based baseline methods and also achieves better performance in most cases. It is worth noting that the two quantitative-based methods both perform poorly on BGL. This may be explained by the fact that the test set of BGL contains many unseen events that never appeared in the training set. Quantitative-based methods rely only on the quantitative patterns among log events in the sequences and fail to capture the semantics in log messages, preventing them from learning more discriminative characteristics to classify the anomalies. This conjecture can also be confirmed by the fact that all sequence- and graph-based methods exploiting semantics in log events work well on the BGL dataset. Another noting thing is that traditional ML-based methods, LR and SVM, perform even better than some sequence-based deep learning methods on the Spirit and Thunderbird datasets especially under smaller window settings although they are still suboptimal than LogGD in most cases. It implies that the quantitative pattern among log events as an important indicator should not be neglected in log anomaly detection. In addition, the overall performances in the F1 score metric of both NeuralLog and LogGD show better than the other sequence-based methods like CNN and LogRobust. This can be explained as Transformer-based methods, including NeuralLog and LogGD, both benefit from exploiting the extra information (positional encodings and spacial structure encodings) in the sequence, leading to better and stable performance. Finally, another thing that should not be overlooked is that all the methods have a trend of significant growth in detection performance on the Thunderbird dataset as the data window size decreases. The reason for this may be the increase in the number of anomalies due to the reduced window size, which enables all the methods to improve the detection performance with the further help of oversampling.

\begin{figure*}[!htb]
  \includegraphics[width=0.3\linewidth]{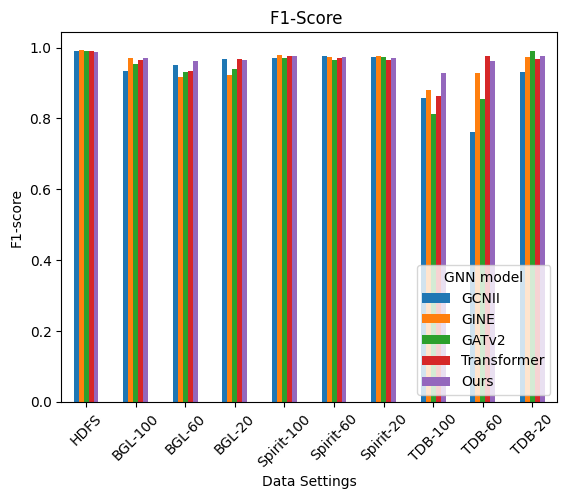} \quad
  \includegraphics[width=0.3\linewidth]{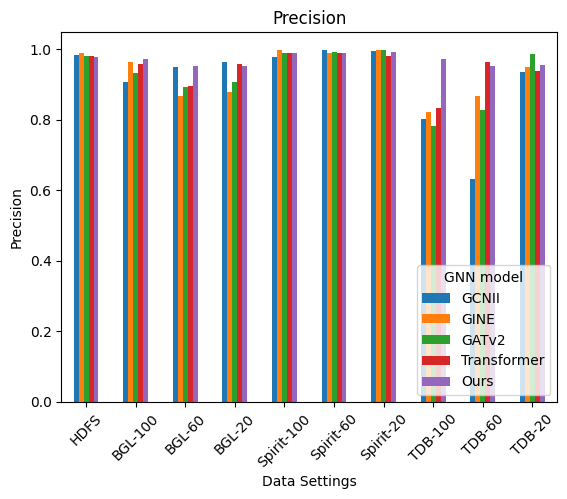} \quad
  \includegraphics[width=0.3\linewidth]{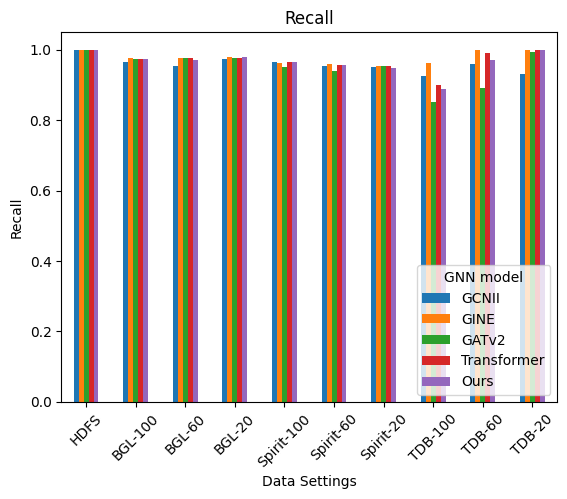}
\caption{Performance Comparison with Different GNN Models}\label{fig:gnn_comparision}
\end{figure*}

\textbf{RQ3. How does LogGD perform with other GNN models?}

In this experiment, we investigate the impact of using different types of GNN layers for graph representation learning on the overall detection performance, i.e., the ablation experiment with GNN models. 
We replaced our customized graph transformer layer with GCNII~\cite{chen2020simple}, GINE~\cite{hu2019strategies}, GATv2~\cite{brody2021attentive}, and TransformerConv~\cite{shi2020masked} models. We did not compare with GCN~\cite{kipf2016semi}, GIN~\cite{xu2018powerful}, and GAT~\cite{velivckovic2017graph} because GCNII, GINE and GATv2 have shown better performance than the corresponding counterparts in their studies. We present the results in Precision, Recall, and F1-Score on the four datasets under the session window setting(HDFS only) and the 100logs, 60logs and 20logs window setting in Fig~\ref{fig:gnn_comparision}. As seen from the figure, all variants of LogGD achieve good performance on the four datasets, while LogGD with our customized graph transformer layer slightly outperforms the variants with other GNN layers in most cases. Furthermore, both LogGD with customized graph transformer layer and the variants with the TransformerConv layer outperform the other variants. 
The results show that the Transformer architecture can overcome the limitations of other Message Passing Neural Network (MPNN) models and better utilize graph structure attributes to learn the normal and anomalous patterns for graph classification.

\textbf{RQ4. How do the specific structural features and the interaction affect the performance of LogGD?}

\begin{figure*}[!htb]
  \includegraphics[width=0.3\linewidth]{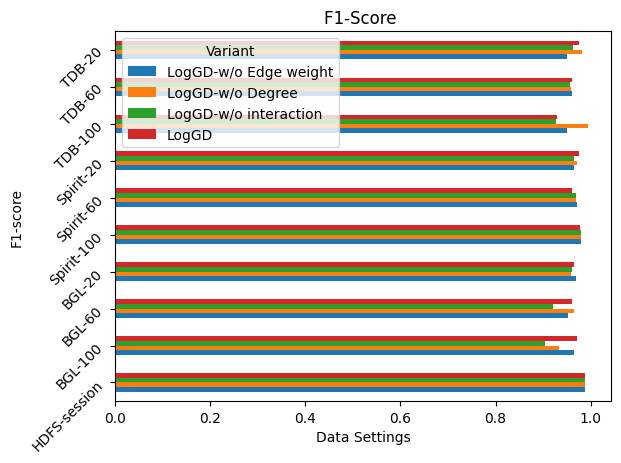} \quad
  \includegraphics[width=0.3\linewidth]{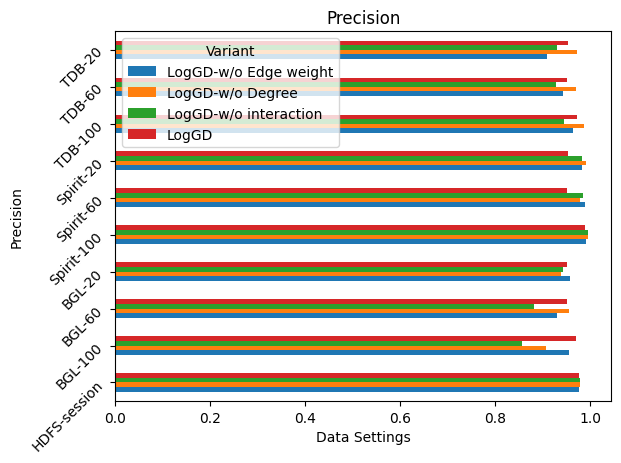} \quad
  \includegraphics[width=0.3\linewidth]{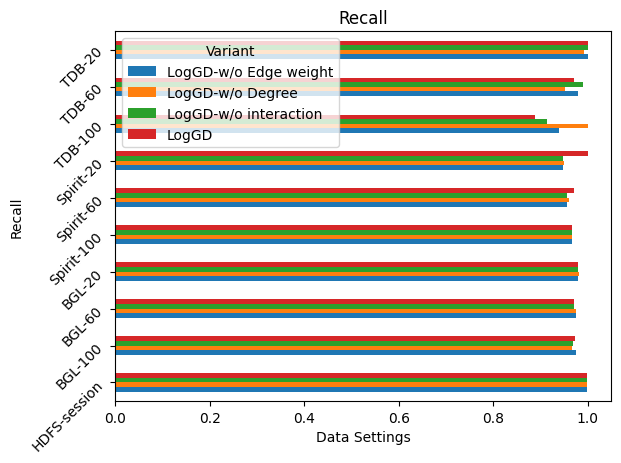}
\caption{Ablation Experiment With Input Feature}\label{fig:feature_comparision}
\end{figure*}

In this experiment, we aim to investigate the impact of graph representations with/without specific structural attributes and with/without the interaction between node features and graph structure on the overall detection performance, i.e., the ablation experiment with the input features. We present the results in Precision, Recall, and F1-Score on the four datasets in Fig~\ref{fig:feature_comparision}. From the figure, we can see that all variants of LogGD achieve good performance with an over 90\% F1 score on the four datasets under different window settings. Second, the effect of the structural attribute and the interaction between node features and structure on performance appears to be data-dependent. For example, the performance of almost all variants with/without a specific attribute is slightly different on the HDFS and Spirit datasets. In contrast, the variation in performance seems to be higher on BGL and TDB datasets for variants that include/exclude the corresponding attribute. Furthermore, we can see that the performance of the variant approach without the interaction between node features and structure (i.e., the distance embedding is directly added to the attention coefficients) degrades significantly on the BGL and Thunderbird datasets. This may indicate that the interaction between node features and structure does help to improve the model's discriminative power between abnormal and normal. In contrast, the effects of the degree alone and edge weight alone appear to be less significant. Interestingly, the variant that excludes the degree attribute performs better on the Thunderbird dataset. In future work, we will further investigate how to better utilize the degree attribute and how to better represent the node-centered topology structure to improve detection performance. Finally, we can see that LogGD, which combines three graph structure attributes, including the degree,  distance and edge weights, and the interactions between node features and structure always perform better in most data settings. This confirms the advantages of combining graph structure and node features.

\section{Discussion}
\label{section:discussion}
\subsection{The advantages and limitations of LogGD}
Two main reasons make LogGD perform better than the related approaches. First, LogGD can capture more expressive structure information from graphs than purely sequential relations between log events. These enhanced features help LogGD better identify the anomalous log sequences. Second, the customized graph transformer model captures the interaction between node features and graph structure represented by the shortest relative path distance, which can also improve anomaly detection performance. 

Our study demonstrates the effectiveness of LogGD for anomaly detection. However, LogGD still has limitations. Our method is a supervised method, which means that intensive data labeling work is inevitable, which may limit the adoption of this method in the industry. In our future work, we will consider self-supervised techniques to enable LogGD to work in semi-supervised or unsupervised information modes to improve the adaptability of the proposed method in real-world scenarios. Also, LogGD identifies anomalies at the graph level. Developers and operators may have to inspect each event in the data window to locate the potential fault~\cite{landauer2022deep}. It would be interesting to explore the feasibility of more fine-grained anomaly detection to reduce the effort and time to locate a fault.

\subsection{Threats To Validity}
In this study, we identify the following threats to validity:

1) External validity threats: this threat denotes those factors that affect the generalization of results. In this study, the external threat to validity lies in the selected datasets, i.e., subject selection bias. In our experiment, we only use four log datasets as experimental subjects. However, these systems are from real-world industrial systems and are widely used in the existing work \cite{he2016experience,meng2019loganomaly,zhang2020anomaly, le2022log}. We believe that 
these four log datasets are representative. In the future, we will evaluate LogGD on more datasets and systems.

2) Internal validity threats: this refers to threats that may have affected the results and have not been properly considered. In this study, the internal threat to validity mainly lies in the implementations of LogGD and compared approaches and the design of the experiments. To reduce the threat from the implementation, we inspect and test the program carefully and assume that all the bugs revealed by testing are fixed in our approach. We implemented LogGD based on popular libraries. 
Regarding the compared methods, we utilize open-source implementations of these methods. For the threat posed by the experimental design, on the one hand, we ran all experiments three times and reported the average of the results. On the other hand, we compare all methods under the same setting for all experiment settings. For example, all the deep learning-based methods share the semantic extraction scheme, the oversampling scheme, and the early stopping scheme. We believe our experimental design can yield fair comparisons among all evaluation methods.

\section{Related Work}
\label{section:related_work}

\subsection{Graph Positional Encoding}
Positional Encoding(PE) is commonly used in image- and text-based tasks with deep learning models, such as image classification~\cite{chu2021conditional} and language translation~\cite{shaw2018self}. 
It plays a crucial role in improving model effectiveness. Studies~\cite{dwivedi2021graph, wan2021glad} have shown that PE representing the structural attributes of graphs is also essential for prediction tasks on graphs. However, finding such positional encodings on graphs for nodes is challenging due to the invariance of graphs to node permutation. Existing PE schemes can be classified into index PE, spectral PE, diffusion-based PE and distance-based PE. For index PE, one option is that indices based on a preset of rules are assigned as positional encodings to nodes in the graph, such as ~\cite{wan2021glad}. However, this scheme still follows a sequence pattern and does not fully exploit the spatial structural attributes of graphs. Another way to build an index PE is to use $n!$ possible index permutations or sample them to train the model~\cite{murphy2019relational}. However, this may result in either expensive computations or the loss of precise positions. Regarding spectral PE, it uses Laplacian Eigenvectors as a meaningful local coordinate system to conserve the global graph structure. However, spectral PE suffers from sign ambiguity, which requires random sign flipping during training for the network to learn the invariance~\cite{dwivedi2021graph}. Diffusion-based PE, such as~\cite{dwivedi2021graph, mialon2021graphit}, is based on the diffusion process, such as Random Walk and PageRank. However, it turns out that they tend to be dataset dependent~\cite{rampavsek2022recipe}. Recently, the shortest path distance has been used as positional encoding in~\cite{ying2021transformers,park2022grpe} and shows promising results. In this work, we utilize the shortest path distance as the basis of graph global structure encoding.

\subsection{Log-based Anomaly Detection}

Log-based anomaly detection has been intensively studied in recent decades. Existing log anomaly detection approaches can be roughly categorized into quantitative-based, sequence-based and graph-based methods in terms of the input data. 

Quantitative-based methods work on log event count matrix. Generally, they can be further divided into traditional ML-based and invariant relation mining-based methods. Traditional ML-based methods, such as LR~\cite{bodik2010fingerprinting}, SVM~\cite{liang2007failure}, PCA (Principal Component Analysis)~\cite{xu2009detecting} and LogCluster~\cite{lin2016log}, are often more efficient compared with deep learning based methods in terms of time costs. Invariant relation mining-based methods, such as Invariants Mining~\cite{lou2010mining}, ADR~\cite{zhang2020anomaly} and LogDP~\cite{xie2021logdp}, have the advantages of low labeling cost and interpretability because they usually work in semi-supervised or unsupervised mode and can capture meaningful relations. Despite these advantages, quantitative-based methods tend to suffer from unstable performance in some specific cases because they cannot capture sequential patterns and semantic information between log events.

In contrast, sequence-based methods take a sequence of log events as input. They typically use various deep learning models to learn sequential patterns in log sequences for anomaly detection, showing impressive performance. Such methods include DeepLog~\cite{du2017deeplog}, LogAnomaly~\cite{meng2019loganomaly}, LogRobust~\cite{zhang2019robust}, CNN~\cite{xu2018powerful} and NeuralLog~\cite{le2021log}. A potential weakness is that most sequence-based methods rely only on sequential relationships between log events in log sequences for anomaly detection. The inability to capture more informative structure information from log sequences hinders sequence-based methods from further improving detection performance and exhibits significant performance fluctuations in some specific cases.

In the past decade, Graph Neural Networks (GNNs) have attracted much attention for their successful applications in many areas, such as drug discovery~\cite{jiang2021could}, social network analysis~\cite{min2021stgsn}, network intrusion detection~\cite{zhou2021hierarchical}, and so on. Recently, the authors of GLAD-PAW~\cite{wan2021glad} apply GAT to log anomaly detection and confirm the feasibility of graph-based log anomaly detection methods. However, although their approach is graph-based, their structure-aware design still follows a sequence pattern, i.e., using only positional encoding, fails to fully exploit the spatial structure of graphs that combines local structure, global structure, and quantitative relationships between nodes. In addition, GLAD-PAW does not consider the interaction between node features and structure of the graph, which may lead to a sub-optimal result. As a graph-based method, LogGD utilizes a combination of node-centered local structure, global structure containing node locations, and quantitative relationships of connections between nodes to generate an expressive representation for a given graph. In the representation learning stage, the interaction between node features and graph structure is combined through a customized Graph Transformer Network, which contributes to the improvement and stability of log anomaly detection performance and demonstrates the powerful effectiveness of graph-based methods. 

\section{Conclusion}
\label{section:conclusion}

In this paper, we have proposed a graph-based log anomaly detection method, LogGD, which can detect system anomalies from logs with high accuracy and stable performance  by combining the graph structure and the node features. Our experimental results on four widely-used public datasets illustrate the effectiveness of LogGD. We hope that LogGD can inspire researchers and engineers to explore further the application of graph neural networks in log analysis.

\section*{Acknowledgment}
This research was supported by an Australian Government Research Training Program (RTP) Scholarship, and by the Australian Research Council’s Discovery Projects funding scheme (project DP200102940). The work was also supported with super-computing resources provided by the Phoenix High Powered Computing (HPC) service at the University of Adelaide.




\bibliographystyle{IEEEtran.bst}
\bibliography{reference}
%





\smallskip
\end{document}